# Artificial Intelligent Atomic Force Microscope Enabled by Machine Learning

Boyuan Huang [1], Zhenghao Li [1,2], and Jiangyu Li [1,3,*]

1. Department of Mechanical Engineering, University of Washington, Seattle, WA 98195-2600
2. Department of Mechanical Engineering, Tsinghua University, Beijing 100084, China
3. Shenzhen Key Laboratory of Nanobiomechanics, Shenzhen Institutes of Advanced Technology, Chinese Academy of Sciences, Shenzhen 518055, Guangdong, China

**Abstract**

Artificial intelligence (AI) and machine learning have promised to revolutionize the way we live and work, and one of particularly promising areas for AI is image analysis. Nevertheless, many current AI applications focus on post-processing of data, while in both materials sciences and medicines, it is often critical to respond to the data acquired on the fly. Here we demonstrate an artificial intelligent atomic force microscope (AI-AFM) that is capable of not only pattern recognition and feature identification in ferroelectric materials and electrochemical systems, but can also respond to classification via adaptive experimentation with additional probing at critical domain walls and grain boundaries, all in real time on the fly without human interference. We believe such a strategy empowered by machine learning is applicable to a wide range of instrumentations and broader physical machineries.

---

[*] Author to whom the correspondence should be addressed to: jjli@uw.edu .

Artificial intelligence (AI) and machine learning have promised to revolutionize the way we live and work. In ancient game of Go, AI has shed unprecedented new insights that have not been recognized by mankind over several thousand years;[1] in medicine, AI has offered diagnosis that rivals the best human doctors;[2,3] and in physics as well as materials sciences, AI has enabled accelerated discovery of new substances, compounds, and mechanisms.[4–8] One of particularly promising areas for AI is image analysis,[9,10] where it far outperforms human beings in pattern recognitions, capable of discerning subtle features that are elusive to the naked eyes. Indeed, AI has been demonstrated to be very effective in analyzing data from both microscopy and spectroscopy studies.[6,11–14] Nevertheless, many current AI applications in image analysis focus on post-processing of data,[12,15–17] while in both materials sciences and medicines, especially under time- and environment-sensitive circumstance and at elusive points that are not easy to spot, it is often critical to respond to the data acquired on the fly, for example by acquiring additional data in the critical locations of material interfaces or tumors. It is also highly desirable to intervene in real time with manipulative or therapeutic treatments on the spot. Here we develop a strategy toward this vision, by demonstrating an artificial intelligent atomic force microscope (AI-AFM) that is capable of not only pattern recognition and feature identification, but can also respond to classification via adaptive experimentation with additional probing at critical locations, all in real time on the fly without human interference. We believe such a vision is applicable to a wide range of instrumentations and experimentations, embodying the true spirit of the automation of science.[18,19]

Our work was initially motivated by atomic force microscopy (AFM),[20] which is a powerful tool in probing, elucidating, and manipulating materials and structures at the nanoscale. Yet AFM experimentations are very tedious and heavily rely on users' experiences in recognizing usually elusive underlying processes. Very often important yet subtle information is overlooked by the users while conducting experiments, and insights are only realized during the post data processing afterward, which is often too late - it is virtually impossible to get back to the critical locations again for further probing, such as defects, heterogeneities, and interfaces, where the most interesting physics occur. The sample could be decomposed for example, or the elusive critical points could be lost like a needle in a forest. This is an ideal scenario for AI-AFM we propose, which is capable of not only recognition and classification, but can also follow up with additional probing in real time upon critical features for further insight, saving all the trouble for human users. This vision is schematically shown in **Fig. 1**, consisting of an AI-AFM that feeds scanning data to a machine learning algorithm in real time. The algorithm is pre-trained with data for material classification and feature recognition, and based on a particular class of materials recognized by the AI, additional features that are relevant to the underlying system will be identified on the fly, such as domain walls (DWs) or grain boundaries (GBs), among others. Through control algorithm, the probe will get back to the identified critical feature in real time and carry out further experimentation appropriate for the probed system on the fly, yielding additional data for analyzing the underlying physical processes. We emphasize that the concept of AI-AFM is fundamentally different from that of AFM robot,[21] which also utilizes intensive imaging process, yet still relies on user intervention through an augmented reality system. In AI-AFM, all the sophisticated tasks are accomplished automatically in an artificial intelligent manner without user interference, embodying the true spirit of AI in an AFM.

**Results and Discussion**

To demonstrate the concept, we consider dynamic strain-based scanning probe microscopy (ds-SPM) that is widely used to probe electromechanical coupling at the nanoscale, including piezoresponse force microscopy (PFM) [22,23] and electrochemical strain microscopy (ESM),[24–27] both of which excite samples through a charged conductive probe and measure the corresponding local deformation. The electromechanical coupling is ubiquitous in both natural materials, synthetic devices, and biological systems, such as ferroelectric materials,[23,28] lithium ion batteries,[25–27] and voltage gated ion channels,[29] underpinning a wide range of functionalities in information processing, energy conversion, and biological processes. Despite their vast different microscopic mechanisms, these electromechanical couplings often exhibit themselves in ds-SPM as apparent piezoresponse,[30] and it is quite challenging for users to discern their dominating microscopic origin. A couple of examples are shown in **Fig. 2**, wherein the amplitude and phase mappings of dynamic strain for a typical ferroelectric lead zirconate titanate (PZT) and electrochemical $LiVO_3$ (LVO) are compared. While they have quite different microscopic mechanisms, the mappings closely resemble each other except for some subtle difference: the 180º phase reversal at the interface with much reduced amplitude as observed in PZT is a signature of ferroelectric domain wall, which is not present in LVO. The question is whether we can train a machine learning algorithm that first differentiates ferroelectric domains from non-ferroelectric mappings, and then responds with additional probing necessary at critical locations for further analysis, for example identify DWs in ferroelectrics and GB in electrochemical materials, after which detailed probing relevant to the particular system can be carried out across these important materials interfaces.

180° domains are commonly presented in ferroelectric materials to minimize their free energy,[31] and they usually exhibit much reduced piezoresponse on DWs with 180° phase contrast, as revealed by **Fig. 2(c)**. On the other hand, amplitude and phase behaviors of non-ferroelectric solids, such as electrochemical materials, are usually not well defined as revealed by **Fig. 2(f)**, wherein the phase contrast is smaller than 180°. Based on these observations, we employ a support vector machine (SVM) algorithm [32] to develop a physics-based classifier that is capable of extracting ferroelectric DWs pixel by pixel from the inputted PFM mappings, thereby helping distinguishing ferroelectric materials from electrochemical ones, for which a different algorithm is introduced to extract GBs from AFM topography mappings, as detailed in **Fig. S1** in Supplementary Materials (SM). Note that while popular convolutional neural network (CNN) has achieved remarkable success in the field of image recognition,[10] it can only categorize a whole map but not capable of delineating the exact DWs or GBs of interest. Fully convolutional networks derived from CNN is capable of identifying lattice atoms in raw scanning transmission electron microscopy (STEM) data [12,33] and should be applicable to our problem, but it requires extensive GPUs to facilitate the training process as well as a large amount of training data with DWs or GBs accurately labelled at the level of pixel. Such sophistications are unnecessary for our particular application, since SVM-based AI algorithm needs only a small dataset that can be trained in less than 10s on an ordinary PC, making it widely accessible. On the other hand, we also note that due to subtleties and complexities often exhibited in ferroelectric and electrochemical materials, careful classification based on machine learning beyond simple rule-based analysis is necessary for our AI-AFM system, and we will come back to this point later with illustrating examples.

SVM is one of the most widely used machine learning algorithms in industry and academia,[34] as detailed in SM, with its applications ranging from handwritten digit recognition for postal automation services,[35] E-mail spam filtering,[36] and accelerating discovery of new piezoelectric materials.[4] SVM can be easily trained with a set of labeled examples, each of which consists of a fixed number of features $(x_1, x_2, ..., x_n)$ and a label $y$ illustrating whether it belongs to one of the two categories ($y =1$ or 0). As schematically shown in **Fig. 3(a)**, we first prepare a training dataset for our SVM model, wherein amplitude and phase variations across a morphology interface are used as indicators to classify whether the interface is a ferroelectric DW or not. For each pixel $P_0$ (marked as green star) on PFM maps of ferroelectric $LiNbO_3$ in **Fig. 3(a)**, 6 adjacent pixels (marked as white dots) are picked from a line centered at $P_0$ and parallel to its phase gradient. The distance from these 6 pixels to $P_0$ are fixed and their respective amplitude and phase, 14 features in total, are sufficient to represent the local variance across $P_0$. Such 14 features with a label of the pixel (DW or not) is then fed into the SVM model. Since each map contains 256×256 pixels and thus generates almost the same amount of training data (with the exception on the map border), it turns out that only 5 pairs of ds-SPM maps are sufficient to train the SVM model, making it highly efficient compared to CNN that has to use a whole map as one training example. More details about our training set is presented in SM along with artificially simulated dataset in **Fig. S2**. SVM first represents these training data as points in a hyperspace, the dimensionality of which depends on the number of features *n*. It then tries to find a hyperplane $\mathbf{w}^T \cdot \mathbf{x} + b = 0$ to separate these points into two categories, as schematically shown in **Fig. 3(a)**, where $\mathbf{w}$ is the feature weights vector and $b$ is the bias term.[32] The points that are most close to the hyperplane from both sides are called the support vectors, as marked on **Fig. 3(a)**, and the larger distance from these two points to the hyperplane (so-called functional margin), the better performance of the trained classifier model. Finally, testing data from new images that are denoised by a local median filtering with a 3-by-3-pixel window can be mapped into the same space, and then classified according to which side of the hyperplane they fall into, revealing whether it is on a DW or not. Note that the pixel-by-pixel recognition of SVM makes it possible to adaptively adjust experimental parameters during scanning, while for CNN that relies on full mapping for recognition, such real time adjustment is not possible. This process is repeated for all pixels except those on the border of maps, creating a binary mask with DWs marked as True. Since DWs are continuous lines on the map, the length of longest line on the binary mask is used to judge whether there are ferroelectric DWs or not. One positive example is shown in **Fig. 3(b)**, wherein 180° DWs in molecular crystal of diisopropylammonium bromide has been correctly identified by the SVM, as marked by the red lines and overlaid on the topography. The negative example of **Fig. 3(c)**, on the other hand, shows ESM maps of a $LiV_3O_8$ sample for which GBs are identified instead. In fact, we have surveyed 7174 ds-SPM mappings accumulated in our lab in the past 10 years, and the normalized confusion matrix in **Fig. 3(d)** shows that 97.3% of 475 maps predicted having 180º DWs are correctly classified, while 99.6% of 6699 maps predicted having no 180º DWs are properly identified as well. Additional examples are presented in **Fig. S3**. These results confirmed that the SVM-based AI algorithm is capable of classification and feature identification of ferroelectric materials with 180º DWs. On the other hand, popular Canny edge detector often fails, as detailed in **Fig. S4**. This is because phase contrast at ferroelectric DWs in real materials often deviates from idealized 180º, and it is often interfered by topography features as well. Since edge detector are highly sensitive to the gradient of a map, slight phase distortion due to scanning disturbance, impurities, topography variation, or artificial pattern may cause false identification, as made evident in **Fig. S4**, while our machine learning algorithm does not suffer from such problems. This demonstrates the need for machine learning-based AI instead of simple rule based analysis in the classification and identification.

Our ultimate goal is to demonstrate an AI-AFM platform that integrates the SVM algorithm with AFM control that is capable of not only classifying ferroelectric materials with 180º DWs in real time, but also adopt adaptive experimentation on the fly to probe the characteristics and mechanisms of apparent piezoresponse in details at critical material interface, i.e. DWs in ferroelectrics and GBs in electrochemical materials. Such additional probing is necessary to confirm the classification without ambiguity, a common practice by human users. To this end, two blind experiments have been conducted on ferroelectric PMN-Pt single crystal and electrochemical Ceria ceramic, as detailed in SM, both of which unknown to the AI-AFM in advance, and the screenshot of the complete experimental processes are recorded. As is shown in **Mov. S1**, PMN-Pt was determined to be ferroelectric having 180º DWs during scanning, and representative amplitude, phase, and topography (overlaid with identified DWs) mappings are shown in **Fig. 4(a-b)**. After such preliminary classification a "ferroelectric routine" was triggered, with the scanning probe zoomed in on an identified DW and carrying out switching spectroscopy PFM (SS-PFM) experiments [37] on a line of points across DW, yielding hysteresis and butterfly loops of **Fig. 4(c)** characteristic for ferroelectric materials, and thus confirm the ferroelectric classification without ambiguity. When Ceria was tested, as shown in **Mov. S2**, the AI-AFM found no 180º DWs from its amplitude and phase mappings in **Fig. 4(d)**, and thus a "non-ferroelectric routine" was triggered to identify GBs overlaid on topography in **Fig. 4(e)**, after which the scanning probe was zoomed in on an identified GB and carried out measurements of first and second harmonic piezoresponses [38,39] across GBs. As seen in **Fig. 4(f)**, second harmonic piezoresponse dominates the first harmonic one in Ceria, characteristic of electrochemical materials, and thus confirm its non-ferroelectric nature without ambiguity. As is clear from the movies, both experiments have been conducted in artificial intelligent manner without human users' interference, demonstrating the capability of our AI-AFM system. Note that the probed sample could be neither ferroelectric nor electrochemical, which can be revealed by the characteristics of first and second harmonic piezoresponse across grain boundaries.

What we demonstrate here is a simple yet powerful prototype artificial intelligent AFM that is trained to carry out complicated scientific experiments from beginning to end, all on its own, and it is just matter of time to incorporate more profound physical processes and more sophisticated deep learning algorithm to expand its power. We expect that similar strategy can also be developed for a wide range of scientific instruments from transmission electron microscope to X-ray diffractometer, as well as a broader physical machineries and systems that heavily rely on human experience to operate at the moment. It is also conceivable that an eco-system can emerge from such vision that all the AI-AFM are interconnected to share and strengthen training data, machine learning algorithm, as well as control, so that user experiences and know-how are no longer limited in a particular lab, but readily spread over the network, and we have made our algorithms publicly accessible to facilitate such movement.[40] More importantly, a general intelligent machine may evolve from such eco-system, which is capable of all round AFM experiments and analysis on its own, revolutionizing the way we do AFM experiments.

**Methods**

**Extraction of grain boundaries (GBs).** As shown in the upright inset of **Fig. S1(a)**, the first-order derivative of height map usually is not continuous at GBs, leading to an overwhelming second-order derivative values there compared to those of nearby area. By setting a threshold value at 90% of the maximum second-order derivative of the map, a binary mask of GBs can be produced as shown in **Fig. S1(d)**.

**Training data**. Considering that the histograms of phase maps are always concentrated around several specific angles, delivering very sparse information, we randomly change the phase offset when picking the 14 features to build a robust model that can work for other realistic cases with various phase offsets. We also intentionally vary the scaling of amplitude features and add white noises to them for the same reason. Although real maps with manually highlighted DWs can be utilized as training dataset, we also succeed in training a model based on simulated maps without tedious labelling work. Specifically, a binary mask of random DWs is first generated and then rendered to mimic the pattern of real maps with respect of morphology, as shown in **Fig. S2**.

**Implementation of Support Vector Machine (SVM).** Functions "fitcsvm" and "predict" in MATLAB are employed to train SVM models and classify new examples, separately. Gaussian kernel is used to implicitly map input dataset into high-dimensional feature spaces,[41] enabling an efficient non-linear classification. Since amplitude and phase values are two set of independent features, we use them separately to train two different SVM models – Amp model and Pha model. When doing prediction, the algorithm will assign a pixel as point of DWs if and only if both Amp and Pha models determine it to be DW based on input features, as shown in **Fig. S3**.

**Canny edge detector.** It is a technique to extract edges from different vision objects in images.[42] In this work, we used edge function in MATLAB with canny method and default setting.

**First and Second Harmonic.** The first and second harmonic PFM responses originates from piezoresponse and other electromechanical mechanisms.[38] For each measured point, a set of AC voltages from 1.5 to 7.5 V were applied with an increment of 0.5 V. At each voltage step, the sample is excited around $f_0$ first and then around $f_0/2$, thereby generating two set of tuning data around $f_0$. The corresponding first and second harmonic amplitude can be extracted by fitting the raw data with the SHO model.

**AI-AFM system.** The AI-AFM experimentation is performed with a commercial Asylum Research MFP-3D AFM. Nanosensors PPP-EFM conductive probes were used for all data shown in **Fig. 4**. The system first conducts a DART PFM mapping to survey possible DWs with the pretrained SVM model. If DWs are found, the probe will move to locate in the middle of the longest DW identified. Then, the system will zoom in on this area with a scan size half of previous one to double-check those DWs. Finally, the middle point on the DW as well as other 4 points across it will be marked so that a set of SS-PFM tests can be completed on each of these points. If no DWs are found, the material will be further assessed for its apparent piezoresponse mechanism. The system will highlight GBs in a height map by using the method discussed in **Fig. S1** and then zoom in on a specific GB to finish a few first and second harmonic comparison experiments. The decision-making process of location here is similar to that of DWs.

Corresponding video can be seen in Mov. S1 and S2. The AI is implemented with MATLAB code and can automatically send commands to AFM after analyzing the scanning data on the fly.

**Acknowledgement**

Supports from National Key Research and Development Program of China (2016YFA0201001), National Natural Science Foundation of China (11627801 and 11472236), and National Science Foundation (CBET 1435968) are acknowledged.

**Availability of data and codes**

All the data and algorithms of AI-AFM used in this paper can be requested from authors.

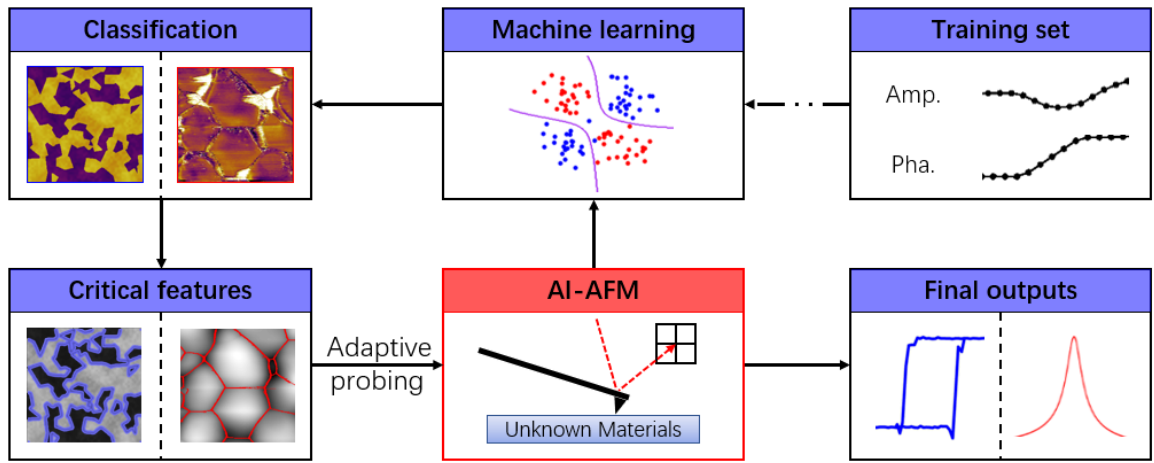

**Fig. 1** The concept of AI-AFM that feeds scanning data to machine learning in real time and classifies samples under probing accordingly, with appropriate features identified. Additional experiments are then carried out on the fly near critical spots for additional data and further insight, all without human interference.

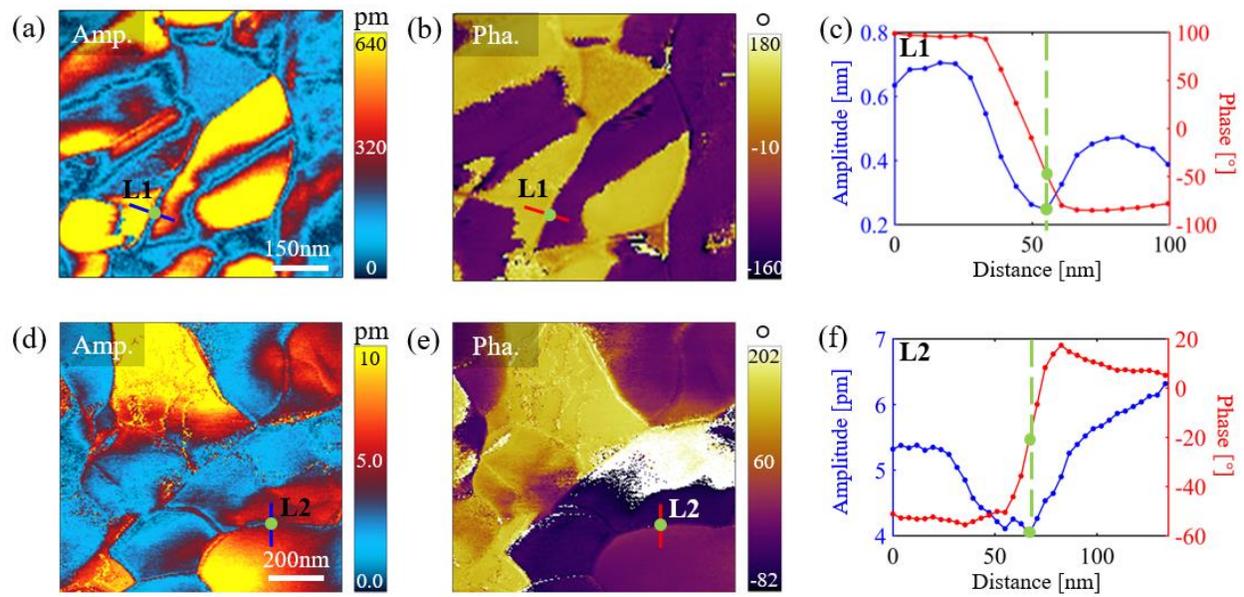

**Fig. 2** Comparing the amplitude and phase mappings of dynamic strain measured by ds-SPM for (a-c) ferroelectric PZT and (d-f) electrochemical LiVO$_3$, along with their respective line profiles.

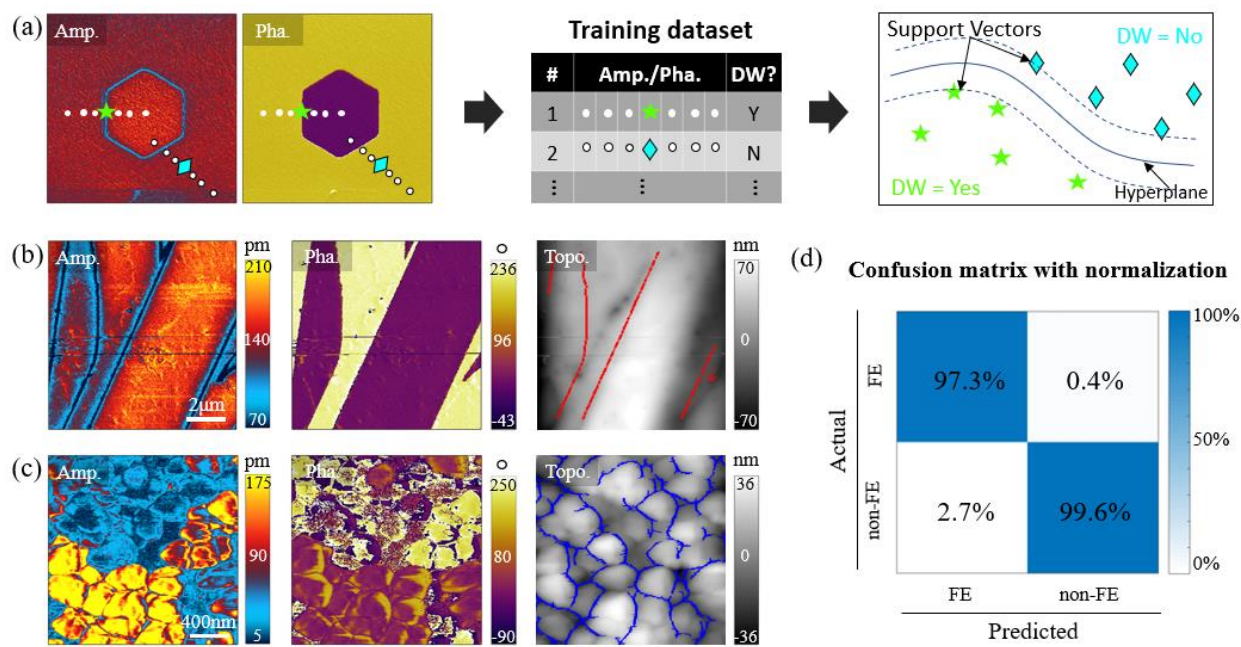

**Fig. 3** Illustration and performance of SVM algorithm for AI-AFM; (a) schematics of training dataset using PFM mappings of LiNbO$_3$ and the corresponding classification of ferroelectric DWs; (b) PFM mapping of ferroelectric diisopropylammonium bromide, with 180° DWs identified and overlaid on topography; (c) ESM mapping of electrochemical LiV$_3$O$_8$, with GBs identified and overlaid on topography; (d) normalized confusion matrix of classification performance on 7174 ds-SPM maps of various materials.

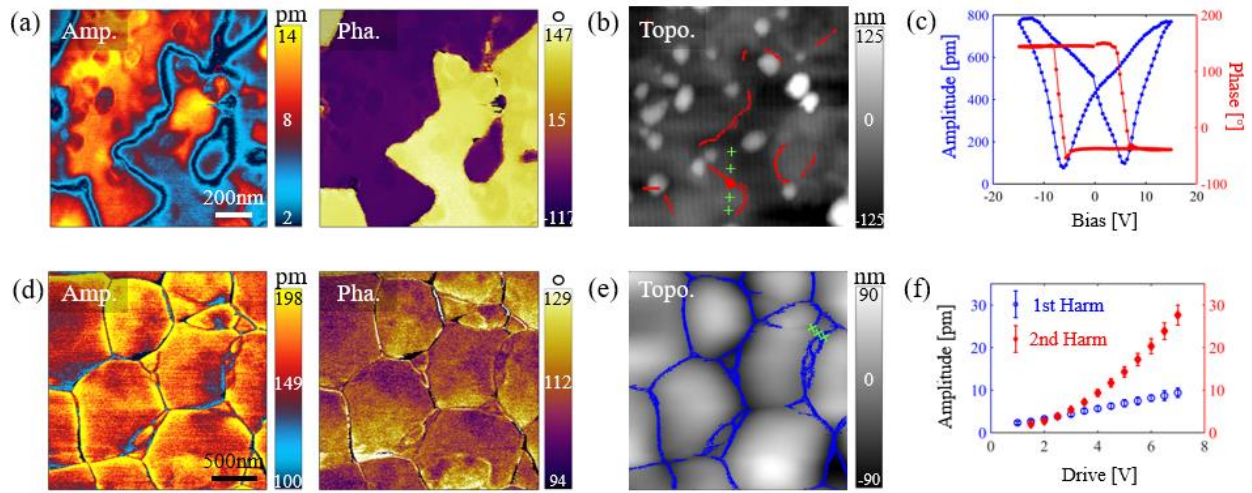

**Fig. 4** Demonstration of AI-AFM for two "unknown" samples that are determined to be ferroelectric (a-c) and electrochemical (d-f); (ad) mappings of amplitude and phase; (be) DWs and GBs recognized; and (cf) ferroelectric hysteresis and butterfly loops on DW and comparison of first and second harmonic piezoresponse on GB, all measured on the fly.

# Artificial Intelligent Atomic Force Microscope Enabled by Machine Learning

## Supplementary Materials

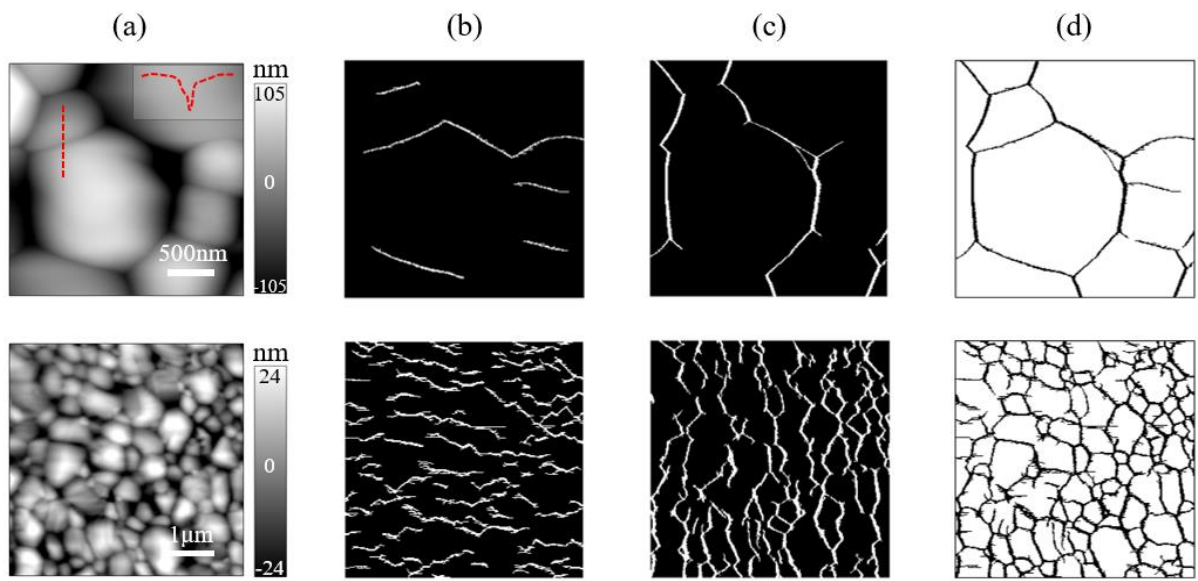

**Fig. S1** GBs of Ceria (1st row) and $CH_3NH_3PbI_3$ (2nd row) identified by the second-order derivative method. (a) Topography and a height profile as an upright inset; (b) second-order derivative along y direction; (c) second-order derivative along x direction; (d) GBs generated from the union of (b-c).

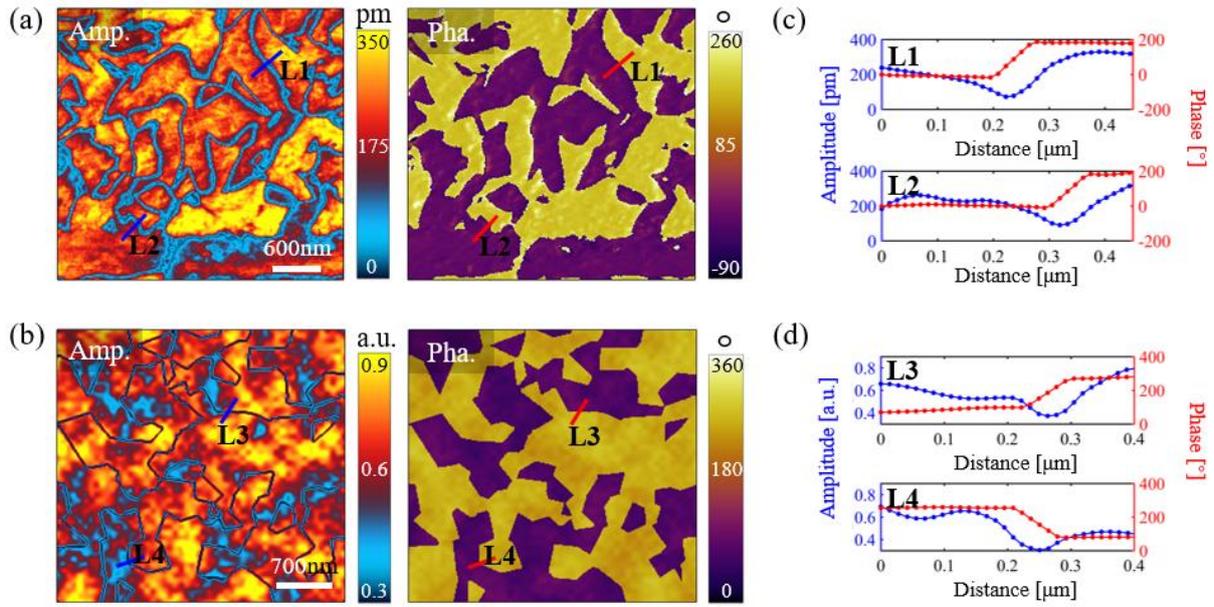

**Fig. S2** (a) Real PFM maps of a PMN-PT sample; (b) simulated PFM maps of a FE sample. Corresponding amplitude and phase scans of: (c) lines L1 and L2, (d) lines L3 and L4.

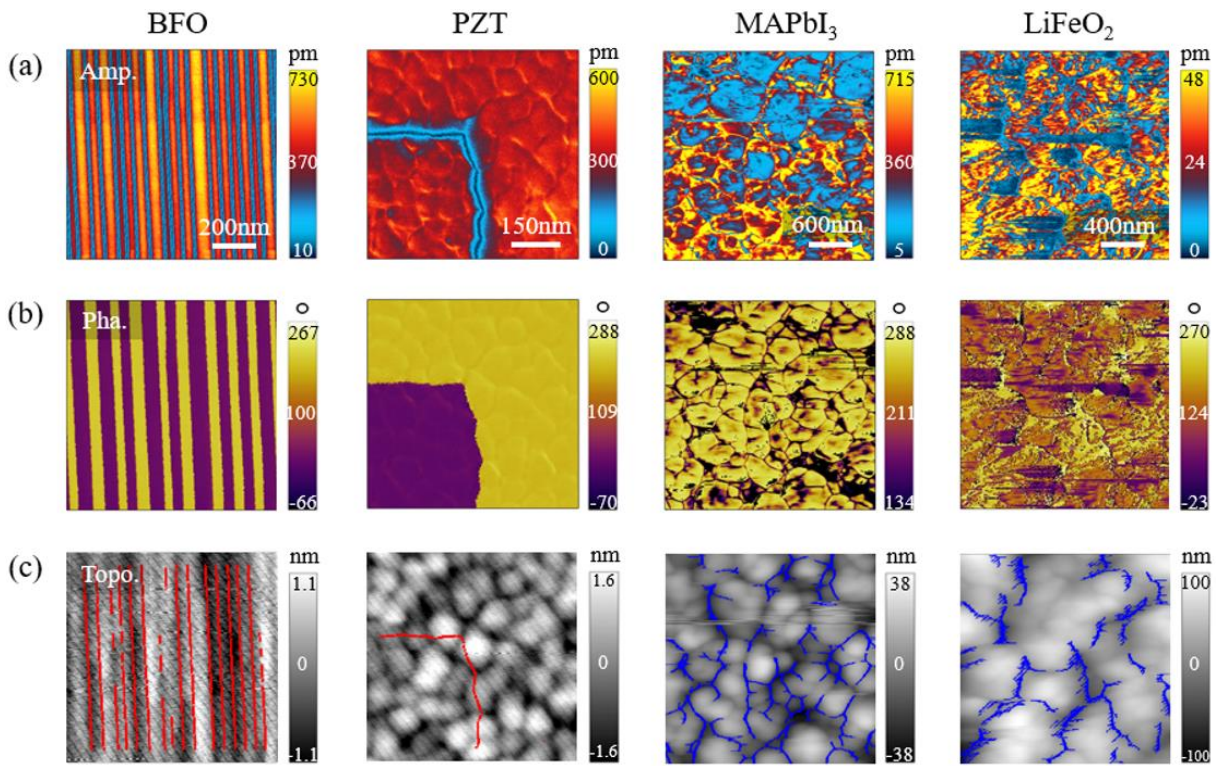

**Fig. S3** Other typical ferroelectric and electrochemical mappings determined by the AI.

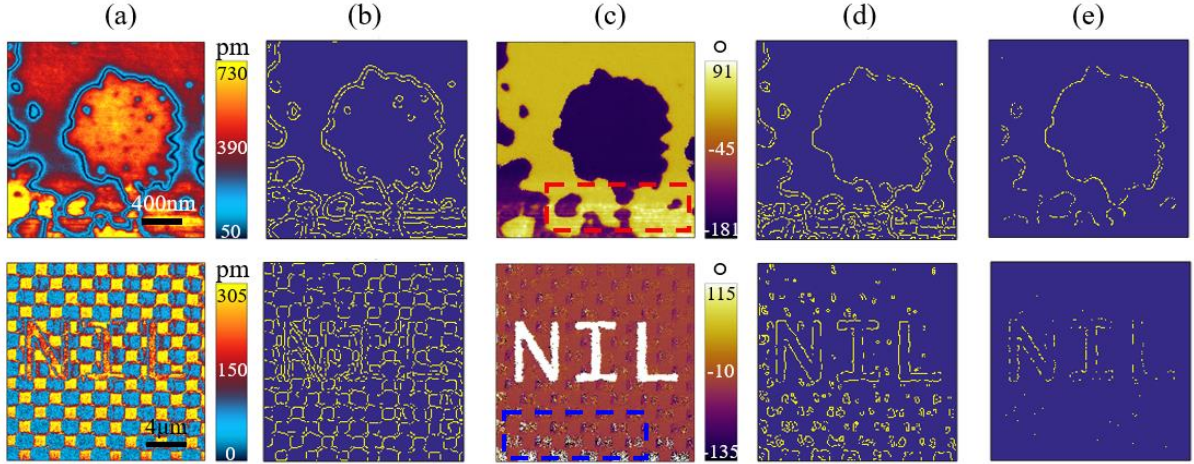

**Fig. S4** Failure of edge-detector algorithm. Comparison between the DWs identified by Canny edge detector and AI algorithms. Real PFM maps are columns (a) Amplitude and (c) Phase, the corresponding edges detected by Canny are shown in columns (b) and (d), separately; (e) DWs identified by AI algorithms.

Even though the detected DWs sometimes could overlap the boundaries of phase maps as in **Fig. 3(b)** and first two columns of **Fig. S3**, it is not always the case. We compare the output of the AI model with a popular Canny edge detector in **Fig. S4**, where columns (b, d) are edges identified by the Canny algorithm from columns (a) Amplitude and (b) Phase maps, separately. Since edge detector are highly sensitive to the gradient of a map, slight phase distortion due to scanning disturbance (marked by red dash rectangular), impurities or artificial pattern (marked by blue dash rectangular) may cause redundant markers, which is not applicable for extracting DWs. In this regard, the DWs highlighted by the AI model will not suffer from such problems and outperform the edge detector, as shown in **Fig. S4(e)**.

**Mov. S1** Movie of AI-AFM testing on a ferroelectric PMN-Pt sample unknown to AI-AFM in advance.

**Mov. S2** Movie of AI-AFM testing on a non-ferroelectric Ceria sample unknown to AI-AFM in advance.